\documentclass[twocolumn,noamsfonts,superscriptaddress,aps,prb,nobibnotes,10pt]{revtex4-1}
\usepackage[paperwidth=208.026mm,paperheight=279.4mm,left=1.5cm,right=1.37cm,top=2cm,bottom=1.72cm]{geometry}

\usepackage[pdftex]{graphicx}
\usepackage[normalem]{ulem}
\usepackage{color}
\usepackage[colorlinks=true]{hyperref}
\hypersetup{
  urlcolor     = blue,    
  linkcolor    = blue,    
  citecolor    = blue     
}
\usepackage[normalem]{ulem}
\newcommand{\etal}{\textit{et al.}}
\usepackage[utf8]{inputenc}
\usepackage{float} 
\usepackage{amssymb}
\usepackage{amsmath}
\usepackage{bm}
\usepackage{xfrac}
\usepackage{ragged2e}
\usepackage{bibunits}
\usepackage{balance}

\begin{document}
\begin{bibunit}[apsrev4-1]

\title{Paramagnon dispersion in $\beta$-FeSe observed by Fe $L$-edge resonant inelastic x-ray scattering}

\author{M. C. Rahn}
\thanks{Present address: Los Alamos National Laboratory, Los Alamos, New~Mexico 87545, USA}
\email[]{rahn@lanl.gov}
\affiliation{Department of Physics, University of Oxford, Clarendon Laboratory, Oxford, OX1 3PU, United Kingdom}
\author{K. Kummer}
\affiliation{European Synchrotron Radiation Facility, Bo\^{i}te Postale 220, F-38043 Grenoble Cedex, France}
\author{N. B. Brookes}
\affiliation{European Synchrotron Radiation Facility, Bo\^{i}te Postale 220, F-38043 Grenoble Cedex, France}
\author{A. A. Haghighirad}
\thanks{Present address: Karlsruher Institut f\"ur Technologie, Institut f\"ur Festk\"orperphysik, Hermann-v.-Helmholtz-Platz 1, D-76344 Eggenstein-Leopoldshafen, Germany}
\affiliation{Department of Physics, University of Oxford, Clarendon Laboratory, Oxford, OX1 3PU, United Kingdom}

\author{K. Gilmore}
\affiliation{European Synchrotron Radiation Facility, Bo\^{i}te Postale 220, F-38043 Grenoble Cedex, France}

\author{A. T. Boothroyd}
\email[]{a.boothroyd@physics.ox.ac.uk}
\affiliation{Department of Physics, University of Oxford, Clarendon Laboratory, Oxford, OX1 3PU, United Kingdom}

\date{\today}

\begin{abstract}
We report an Fe $L$-edge resonant inelastic x-ray scattering (RIXS) study of the unusual superconductor $\beta$-FeSe. The high energy resolution of this RIXS experiment ($\approx$\,55$\,$meV FWHM) made it possible to resolve low-energy excitations of the Fe $3d$ manifold. These include a broad peak which shows dispersive trends between 100 and 200\,meV along the $(\pi,0)$ and $(\pi,\pi)$ directions of the one-Fe square reciprocal lattice, and which can be attributed to paramagnon excitations. The multiband valence state of FeSe is among the most metallic in which such excitations have been discerned by soft x-ray RIXS.
\end{abstract}

\maketitle
\section{Introduction}
Collective magnetic fluctuations that extend up to hundreds of meV in energy feature prominently in the copper oxide and iron-based families of high temperature superconductors~\cite{Dai2015,INO16}. A detailed understanding of these excitations may reveal important information about the corresponding ground states, such as the strength and character of electronic correlations and itinerancy. Moreover, magnetic fluctuations can lead to unconventional forms of superconductivity, and could play a role in the mechanism of copper oxide and iron-based superconductivity~\cite{SCA12,CHU12,Korshunov2018}.

In general, the experimental method of choice to map the spectrum of magnetic fluctuations is inelastic neutron scattering (INS). Modern time-of-flight neutron spectrometers are able to reveal important features of the magnetic spectrum, such as the propagation vector, dispersion, and degree of anisotropy and itinerancy of the magnetism, extending over the entire Brillouin zone with approximately meV energy resolution~\cite{Dai2015,INO16}. The main drawback of INS is the weakness of the neutron's interaction with the sample, which necessitates sample masses on the order of several grams. For single-crystal studies of materials like $\beta$-FeSe, this may require the co-alignment of many hundreds of individual crystallites~\cite{Shamoto2015,Ma2017}.

By contrast, momentum-resolved x-ray spectroscopic measurements at third generation synchrotrons can be performed on crystals with dimensions below 100\,$\mu$m, but face different challenges.  The signal from the relevant excitations may be weak, especially if the scattered beam is to be analysed with an extreme energy resolving power ($E/\Delta E \approx 14,000$ in the present study), and may be complicated by other processes. In the soft x-ray regime, the maximum momentum transfer accessible may not be sufficient to probe the entire Brillouin zone, and the attenuation of the beam adds to the difficulties, requiring ultra-high vacuum (UHV) conditions.

Recent advances in soft x-ray resonant inelastic x-ray scattering (RIXS) instrumentation have significantly improved the combined energy resolution, and now allow a continuous variation of scattering angle without breaking UHV conditons~\cite{Brookes2018}. In combination with the extreme brilliance of focused undulator beams, the soft-x-ray RIXS technique provides a complementary probe to INS for studies of the low-energy collective dynamics at the heart of correlated electron phenomena~\cite{Ament2011}.

\begin{figure}
\includegraphics[width=0.95\columnwidth,trim= 0pt 0pt 0pt 0pt, clip]{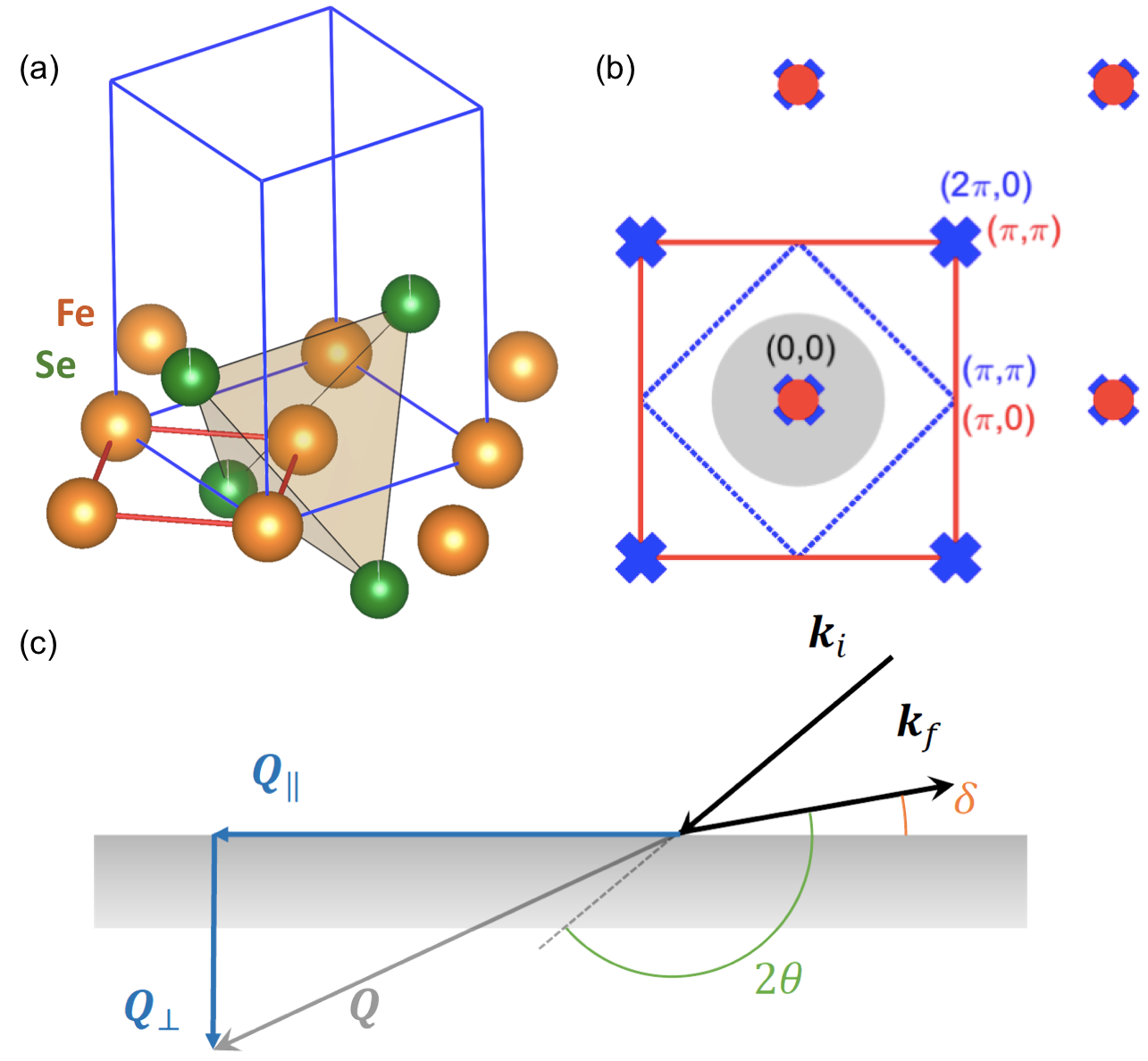}
\caption{\label{Fig1} (a) Structure of FeSe antifluorite layers with indications of the room temperature tetragonal (blue) unit cell (space group $P4/nmm$, lattice parameters $a = 0.376$\,nm, $c =  0.551$\,nm), and the one-Fe (red) unit cell of the square Fe-only lattice ($a = 0.265$\,nm). (b) Two-dimensional view of reciprocal space, showing the tetragonal and one-Fe first Brillouin zones [colors corresponding to panel (a)]. The regime accessible to Fe $L$-edge RIXS is shaded gray. (c) Illustration of the scattering geometry of the present experiment. The projection of the momentum transfer onto the $(HK0)$ plane, $\mathbf{Q}_{\parallel}$, is maximized by choosing a large scattering angle $2\theta$ and a low grazing angle $\delta$. The choice of grazing scattered, rather than grazing incident, geometry minimizes the quasi-elastic background.}
\end{figure}

$\beta$-FeSe (below ``FeSe'') remains one of the most puzzling cases among iron-based superconductors (IBSCs)~\cite{Coldea2018,Boehmer2018}. Structurally, FeSe is the simplest member in the IBSC family, featuring the basic antifluorite layer motif, i.e.~square layers of Fe atoms that are tetrahedrally coordinated by a pnictogen or chalcogen, see Fig.~\ref{Fig1}(a). However, the behavior of FeSe deviates from the typical behavior of IBSCs. The pure material enters a superconducting state ($T_c=9\,$K) without chemical doping, yet $T_c$ is highly sensitive to other tuning parameters, rising to $37$\,K under pressure~\cite{MED09}, $50$\,K by intercalation~\cite{GUO10}, and up to $100$\,K in monolayers~\cite{WAN12,Ge2014}.

Furthermore, whereas typical IBSCs develop superconductivity on suppression of a coexisting antiferromagnetic (AFM) and structurally-distorted nematic phase, in FeSe the nematic phase coexists with superconductivity, albeit on distinct temperature scales, ($T_c = 9$\,K) $\ll$ ($T_s=90$\,K). AFM order is absent from the nematic phase of FeSe, but strong cooperative paramagnetic spin fluctuations are observed~\cite{Rahn2015,Wang2016}. Very recently, quasiparticle interference studies have revealed that the unusual superconducting gap configuration in FeSe favors pair formation only between $d_{yz}$ orbitals~\cite{Sprau2017}. In effect, the ground state of FeSe appears to be shaped by competing structural, electronic and magnetic degrees of freedom. The question remains which (or rather, which combination) of the associated fluctuations (lattice, orbital or magnetic) is driving the strange superconducting state in this material~\cite{Chubukov2015,Glasbrenner15}.

As an advanced technique with the potential to probe both orbital and magnetic correlations, RIXS is particularly promising for disentangling the relevant interactions of this highly complex ground state. Direct RIXS involves the observation of a net energy transfer after a two-step process in which a core-hole is first created and then recombines with an electron from the valence shell~\cite{Ament2011}. This contributes to the x-ray scattering amplitude in second order perturbation theory, which implies a strong resonant enhancement when the incident energy coincides with a core-level binding energy~\cite{Schuelke2007}. In the context of unconventional superconductivity, soft x-ray RIXS has proven particularly successful in transition metal $L_{2,3}$ edge studies, where $2p$ core electrons are being excited to the $d$-electron valence bands~\cite{Dean2015}.

In the strongly correlated $3d^9$ magnetic valence state of cuprate superconductors and their parent compounds, RIXS spectra show a sizable response from the dynamic magnetic susceptibility~\cite{Ament2009}, in excellent agreement with neutron inelastic data~\cite{Braicovich2009,LeTacon2013}. In these copper-based materials, superconductivity arises in the vicinity of archetypal localized magnetism, due to one hole confined to a single ($d_{x^2-y^2}$) orbital. By contrast, for the unconventional superconducting states in iron-based materials, all five $d$ electronic orbitals are generally considered relevant. Consequently, the valence state of IBSCs is more delocalized and less correlated than in cuprates~\cite{Yang2009}. The associated magnetism is described as an itinerant phenomenon, arising from the nesting of hole and electron Fermi pockets~\cite{Chubukov2008}.

This itinerancy has been impeding x-ray spectroscopic studies of electronic correlations in IBSCs, since the dominant fluorescent RIXS response of the Fermi liquid makes it difficult to discern weak excitations of local degrees of freedom~\cite{Gretarsson2015}. Nevertheless, in principle RIXS promises much more information about the complex multi-band nematic magnetic states of IBSCs than neutron spectroscopy, which simply couples to the dynamic magnetic susceptibility via the magnetic dipole interaction. RIXS should be sensitive to any inter- and intra-orbital electronic dynamics, whether involving spin-flips or not~\cite{Kaneshita2011,Nomura2016}. In this regard, the ``local \textit{vs} itinerant'' dichotomy of IBSCs is both promising and problematic. Previously, RIXS in IBSCs has been restricted to the Fe $K$ edge ($\approx7.1$\,keV)~\cite{Jarrige2010,Jarrige2012}, to large energy transfers at the Fe $L$ edge~\cite{Yang2009,Hancock2010,Chen2012,Monney2013,Nomura2016}, or to atypically insulating IBSCs~\cite{Gretarsson2015}. More recently, $L_3$-edge RIXS studies have succeeded in resolving magnon excitations in IBSC parent compounds, such as BaFe$_2$As$_2$~\cite{Zhou2013}, NaFeAs~\cite{Pelliciari2016a}, SmFeAsO~\cite{Pelliciari2016b} and EuFe$_2$As$_2$~\cite{Pelliciari2017}. Paramagnon fluctuations in doped systems have only been observed in the two cases of (Ba,K)Fe$_2$As$_2$~\cite{Zhou2013} and Na(Fe,Co)As~\cite{Pelliciari2016a}.

We here report the observation of similar excitations in an iron chalcogenide, FeSe. Using the new ERIXS end-station of beamline ID32 at the European Synchrotron Radiation Facility~\cite{Brookes2018} with an energy resolution of 55\,meV (FWHM) at the Fe $L_3$ edge, it was possible to distinguish weak low-energy dispersive excitations from the (quasi-)elastic peak and fluorescence background. The multi-band valence state of FeSe is among the most metallic systems in which it has been possible to resolve correlated-electron excitations by RIXS. Our observations are qualitatively consistent with cooperative magnetic fluctuations in other IBSCs.

\section{Experiment}
High quality single crystals of tetragonal (space group \textit{P}4/\textit{nmm}~\cite{Margadonna2008}) $\beta$-FeSe were synthesized using a KCl--AlCl$_3$ vapor transport technique~\cite{Chareev2013}. The compound crystallizes in the form of platelets with dimensions on the order of 1\,mm$^2$ and $\approx 200\,\mu$m thickness (typical sample mass $\leq 1$\,mg). We characterized these samples by magnetometry and laboratory x-ray diffraction. As discussed in the Supplemental Material~\cite{supplemental}, these measurements confirmed the high crystalline quality of the samples as well as a sharp onset of the Meissner--Ochsenfeld effect at $T_\textrm{c}\approx 9$\,K.

When exposed to air, FeSe crystals tarnish within days. Immediately before the RIXS measurement, a surface layer was therefore stripped from the crystal platelets using an adhesive tape. This revealed a clean, mirror-like surface. The samples were then directly transferred to the UHV system of ID32. Inside the sample chamber, the copper sample carrier is mounted on the goniometer which is connected to the cold finger of a helium flow cryostat via copper braids. Throughout the experiment the sample stage was held at $\approx$\,21\,K. The beam size at the sample position is $4\times 40\,\mu$m$^2$ (vert.$\times$horz.).

In a compromise for intensity over energy resolution, we configured the spectrometer in its medium-resolution setting, with a combined energy resolution of $\approx55$\,meV FWHM. At the Fe $L_3$ edge (incident energy $E_{\rm i}\approx 707\,$eV, $\lambda\approx1.75\,$nm), we estimate an x-ray penetration on the order of 200\,nm into the sample, corresponding to a depth of several hundred unit cells.

A schematic of the present RIXS scattering geometry, which has been widely used for studies of layered superconductors~\cite{Ament2009,Braicovich2010,Zhou2013}, is given in Fig.~\ref{Fig1}(c). Due to the strongly two-dimensional character of electronic correlations in these materials, a dependence of the spectra on the perpendicular momentum transfer $Q_\perp$ is commonly neglected~\cite{Braicovich2009,LeTacon2013,Pelliciari2017}. For the case of the related material BaFe$_2$As$_2$, the extent of this $Q_\perp$-dispersion is illustrated by the width of the curve drawn in Fig.~\ref{Fig5}(a), as discussed below. The quasi two-dimensional character of FeSe is also directly evident from the cylindrical Fermi surfaces observed by angle-resolved photoemission spectroscopy~\cite{Watson2015}.

The same excitation can therefore be probed at different configurations of scattering and grazing angles ($2\theta$, $\delta$). To enhance the low-energy excitations and reduce the quasielastic peak, spectra at finite momentum transfer were obtained with a grazing scattered beam. The samples were aligned using the specular $(001)$ reflection, which became accessible by increasing the incident energy to 1.7\,keV. Since no other Bragg reflection was accessible as a second reference, the azimuth of the sample was aligned by half-cutting the beam with one edge of the rectangular platelet (which is parallel to a $\langle 100 \rangle$-type direction~\cite{supplemental}). For all measurements, the incoming beam was linearly (``$\pi$'') polarized in the horizontal scattering plane, and the polarization of the scattered beam was not analyzed.

We refer to reciprocal space coordinates in the conventional ``one-Fe'' two-dimensional Brillouin zone, see Fig.~\ref{Fig1}(b). We neglect the orthorhombic distortion and twinning and state coordinates in units of $1/a$ ($a=0.265$\,nm), i.e.~2$\pi$ corresponds to one reciprocal lattice unit~(r.l.u.). Spectra were measured at a number of scattering angles, with the in-plane momentum transfer $Q_{\parallel}$ directed either along the $(1,0)$ or $(1,1)$ direction. As the incident energy is fixed at the Fe $L_3$ edge, the accessible range of scattering angles (50$^\circ$--150$^\circ$) imposes kinematic constraints on the scattering process. For FeSe and related IBSCs, the in-plane momentum transfer is limited to approximately $(0.5,0)\pi$ and $(0.35,0.35)\pi$, as indicated in Fig.~\ref{Fig1}(b).

\begin{figure}
\includegraphics[width=0.95\columnwidth,trim= 0pt 5pt 0pt 0pt, clip]{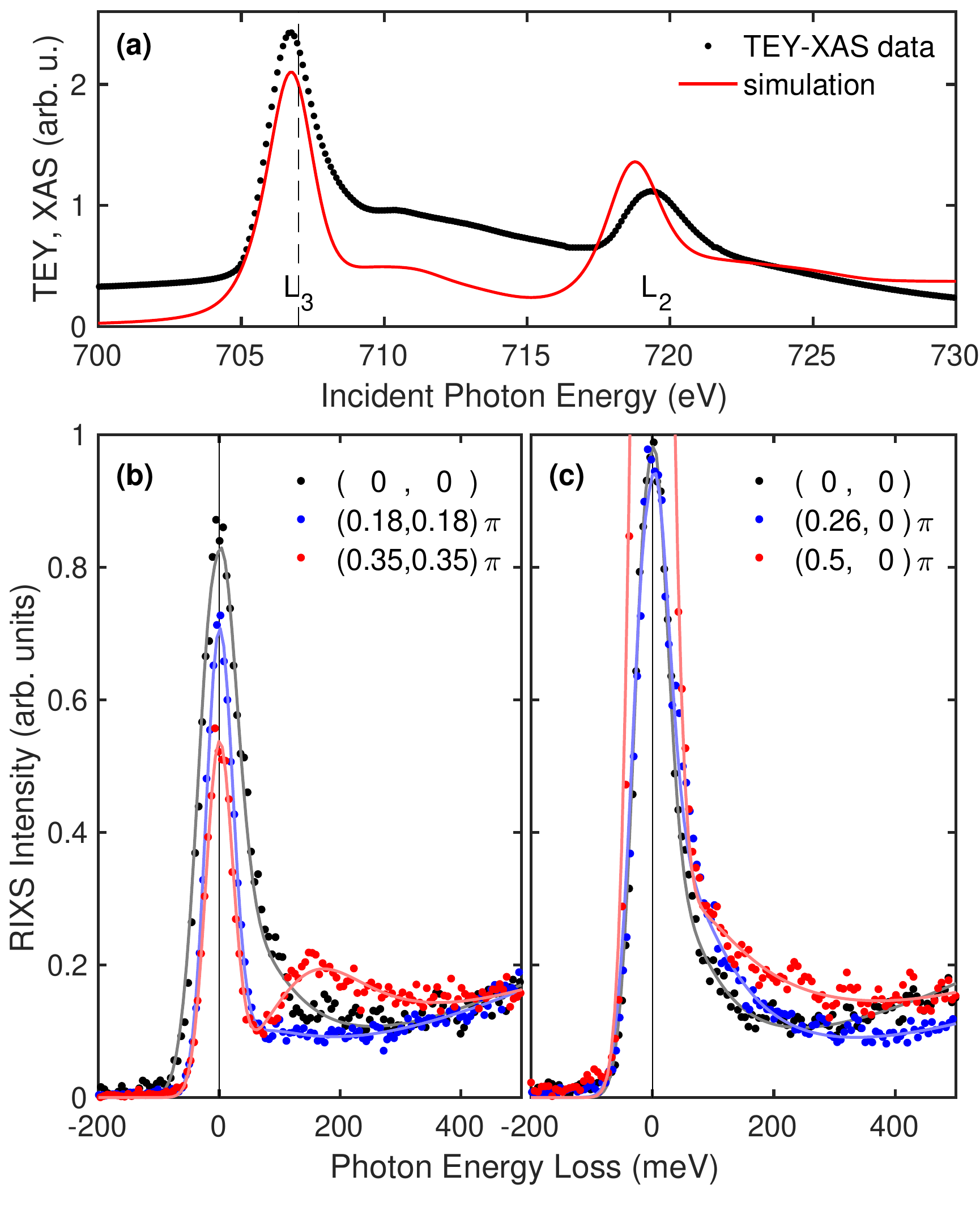}
	\caption{\label{Fig2} (a) Total electron yield (TEY) x-ray absorption spectrum of the Fe $L_{2,3}$ edges in FeSe, measured as the drain current from the sample surface. The simulated absorption spectrum, calculated in the many-body Bethe-Salpeter equation framework, is shown for comparison. The vertical line indicates the photon energy used in the RIXS experiment. (b)--(c) Comparison of RIXS spectra with momentum transfers $Q_\parallel$ along the $(1,1)$ and $(1,0)$ directions. The anomalous low-energy excitations of interest appear well-resolved only for large momentum transfers along (1,1). For other datasets, the local maxima are obscured by a strong overlap with quasielastic scattering and RIXS fluorescence. The solid lines are phenomenological fits to the data (see text for details).}
\end{figure}

\begin{figure*}
\includegraphics[width=1.8\columnwidth,trim= 0pt 0pt 0pt 0pt, clip]{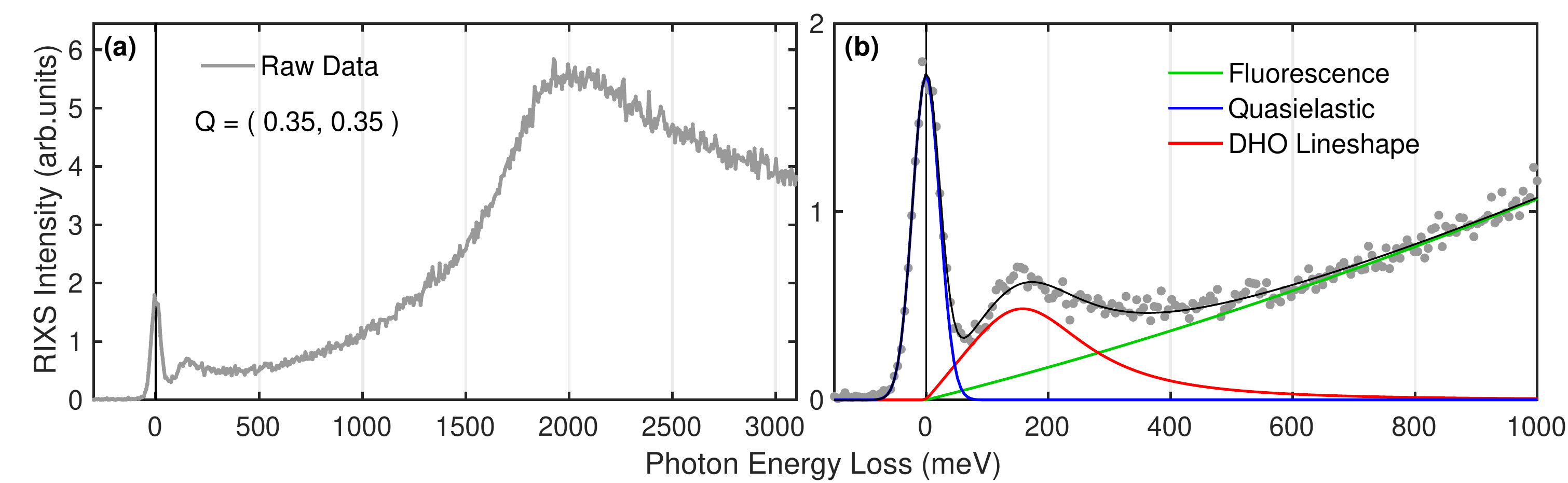}
\caption{\label{Fig3} (a) Raw Fe $L_3$ RIXS spectrum of $\beta$-FeSe recorded at 21\,K, with an in-plane momentum transfer of $Q=(0.35,0.35)\pi/a$. (b) Detailed view of the low-energy region, with lines indicating a least-squares fit of four contributions to the signal.}
\end{figure*}

\section{Results}

\subsection{RIXS response of FeSe}

Figure~\ref{Fig2}\,(a) shows the total electron yield (TEY) x-ray absorption sepectrum (XAS), which is measured as the drain current from the sample surface, over a photon energy range covering the Fe $L_{2}$ and $L_{3}$ edges. The features observed here, in particular the peak shape and shoulder of the $L_3$ edge, resemble those observed in other IBSCs, and also of elemental iron~\cite{Yang2009}. Due to the weakness of the RIXS magnon response and beam time constraints, we were not able to explore in detail the resonant behavior (incident-energy dependence) of the RIXS spectra. The selection of the intermediate state has a characteristic influence on the intensity of inter-band excitations and RIXS fluorescence~\cite{Hancock2010}. We chose to obtain all spectra at $E_i\approx707$\,eV, where the magnon signal is known to be amplified in related materials~\cite{Zhou2013} and the intensity of the sloping fluorescence background at low energy transfers appeared acceptable. The characteristics of the paramagnon spectrum are expected to be independent of $E_i$.

When irradiating the same position on the sample surface for an extended period, we observed a continuous shift of the TEY spectrum toward Fe$^{3+}$ character. To confirm that the finite energy excitations of interest in this study are an intrinsic response of bulk FeSe, we measured a number of reference spectra. In these separate experimental sessions (i.e. using different crystals), the samples were periodically translated in the vertical direction, as to avoid irradiating any point on the sample surface for more than five minutes. The consistency of this data with the original measurements indicates that the signal above $\approx100$\,meV energy transfer is a bulk response, which appears not to be strongly affected by potential surface degradation.

Each RIXS spectrum, summed from a large number of short scans, corresponds to a total data acquisition time of 7--8 hours, at a count rate of on average less than one photon per minute at intensity maxima. To account for variations in resonance strength as a function of incident energy, the data were normalized to the integrated RIXS intensity in the range of $500$--$2800$\,meV photon energy loss.

In Fig.~\ref{Fig2} we show a comparison of raw spectra at different momentum transfers to illustrate the main characteristics of our data. Spectra recorded with momentum transfer $Q_\parallel$ along the $(1,1)$ and $(1,0)$ directions are shown in panels (b) and (c), respectively.

The key interest of our study lies in the weak anomalous contributions that are most intense aroud 80--200\,meV energy transfer. In the present data, these features are only resolved as distinct peaks for large momentum transfers along the $(1,1)$ direction [see $Q_\parallel=(0.35,0.35)\,\pi$ spectrum in Fig.~\ref{Fig2}(b)]. Elsewhere, these features are more difficult to discern, either because they are more strongly damped, or because they overlap with other RIXS contributions.

Close to zero energy transfer, the spectra are dominated by quasielastic scattering. The intensity of this contribution depends on the incident energy, and choice of scattering geometry. 

At higher energy transfers, excitations in the continuum of itinerant charge carriers contribute significantly to the measured intensity. In the present spectra, this RIXS fluorescence contributes a sloping background at low energy transfers and forms a broad maximum at $\approx 1.9$\,eV photon energy loss [see Fig.~\ref{Fig3}(a)]. These characteristics closely resemble measurements of 122 and 1111-type FeAs-based superconductors~\cite{Yang2009,Zhou2013}. In a Fe $L_3$-edge RIXS study of the chalcogenide Fe$_{1.087}$Te, Hancock \textit{et al.} have demonstrated that the slope of the RIXS fluorescence at lower energy losses and its exponential decay above 2\,eV are consistent with the Fermi-liquid response to the creation of the core hole~\cite{Hancock2010}. In the present context, these itinerant excitations are not of direct interest. 

Additional low energy RIXS excitations are barely resolved at the Brillouin zone center, $(0,0)$, where they appear as a shoulder on the quasielastic peak. However, with increasing momentum transfer, we observe a systematic increase in these scattering contributions. They are best resolved along the $(1,1)$ direction, where they form a distinct peak around 150--200\,meV at $Q=(0.35,0.35)\pi$.

\subsection{RIXS calculations}

To rule out the possibility that the feature appearing around 100--200\,meV is due to either electronic orbital excitations or lattice vibrations, we performed first-principles calculations of the electronic contribution to the RIXS spectrum and use a simplified model to estimate the phonon contribution.

The electronic quasiparticle contribution to the RIXS spectrum of FeSe was calculated with the OCEAN code~\cite{Shirley1998,Vinson2011,Gilmore2015} by evaluating the Kramers-Heisenberg equation within the framework of the many-body Bethe-Salpeter equation (BSE).   The BSE treats excitations within a two-particle, electron-hole approximation and includes screened direct and bare exchange Coulomb interactions between the electron and hole by summing ladder diagrams.  The BSE is well-suited for RIXS calculations as it treats both core-conduction (intermediate-state) and valence-conduction (final-state) excitons on equal footing and does not assume that the final-state exciton is restricted to a localized manifold of $3d$ states.  

We first test the accuracy of the BSE method for FeSe by evaluating the Fe-$L_{2,3}$ X-ray absorption spectrum.  The result, shown in Fig.~\ref{Fig2}(a), is largely in agreement with experiment.
The RIXS calculations, shown in Figs. S4(a) and (b) of the Supplemental Material~\cite{supplemental}, reproduce well the fluorescence contribution, showing a broad peak around 2\,eV as observed in the experimental results.  Also in accord with experiment, the calculations show essentially no dispersion of the fluorescence contribution.  Importantly, the calculation of the quasiparticle contribution goes smoothly to zero at zero energy loss, indicating no appreciable intensity associated with distinct $d$--$d$ orbital excitations around 100--200\,meV.

The BSE calculations are based on a density functional theory electronic structure using the local density approximation (LDA) to the exchange correlation functional.  Electronic correlations, beyond those captured by the LDA, can play an important role in the physical properties of iron-based superconductors, including FeSe.  Several quantities, such as effective masses and bandwidths, are not well reproduced by LDA calculations~\cite{Yi2017}.  Computational results are improved by incorporating additional electronic correlations either through the addition of $GW$ self-energies~\cite{Tomczak2012} or by use of dynamical mean field theory (DMFT)~\cite{Aichhorn2010}.  For the XAS calculations presented in \ref{Fig2}(a), we have included a $G_0W_0$ self-energy correction to the LDA electronic structure, which quantitatively improved agreement with the experimental result.  We performed the RIXS calculations at the LDA level after finding that inclusion of the $G_0W_0$ self-energy correction made only small quantitative differences.  While a DMFT electronic structure would yield more narrow bandwidths we do not believe this would result in the appearance of pronounced $d$-$d$ orbital excitations around 100--200\,meV in the RIXS calculations.

We provide a rough estimate of the phonon contribution to the RIXS spectrum using a simple model~\cite{Ament2011a} for the coupling of a localized electronic excitation to a local vibrational mode.  This model includes as parameters the vibrational frequencies, the coupling strength between the electronic excitations and the phonon, and the RIXS intermediate-state core-hole lifetime.  Neutron experiments have shown that vibrational excitations in FeSe are constrained to energies below a sharp cutoff at 40\,meV~\cite{PHE09,Rahn2015}.  Electron–energy-loss spectra reported by Zakeri~\etal~reveal dominant phonon modes centered at 20.5, 25.6 and 40\,meV~\cite{Zakeri2017}. The core-hole lifetime corresponds to that of the Fe $2p_{3/2}$ level, which is known to be 180\,meV~\cite{Krause1979}.  As illustrated in Fig. S4(c) of the Supplemental Material~\cite{supplemental}, the resulting RIXS spectrum due to excitation of phonons is peaked around 40\,meV.  This indicates that single-phonon excitations strongly dominate over the higher-order processes, which contribute a tail of scattering intensity towards higher energies~\cite{Ament2011a,Yavas2010}.  For realistic electron–phonon coupling strengths in FeSe, we can therefore exclude the possibility that the pronounced RIXS excitations above 100\,meV are due to lattice excitations.

It is not possible to calculate the phonon RIXS intensity in absolute units or estimate relative intensities compared to the other spectral features. In the present data, phonon scattering should be best resolved at $Q_\parallel=(0.35,0.35)\pi$, where the anomalous excitation (peaked around 180\,meV) is separated from the quasielastic line [see Fig.~\ref{Fig3}(b)]. As there is no evidence for a peaked feature around 40meV, or any clear shoulder or asymmetry to the elastic peak, we assume that phonon scattering is much weaker than the anomalous (paramagnon) excitation of interest. Nevertheless, we cannot rule out the possibility of a weak phonon contribution which could systematically affect the results of our fits to the spectra with a phenomenological lineshape, as described below.

\subsection{Phenomenological fit}

To illustrate the method used to analyze the anomalous low energy excitations, we display in Fig.~\ref{Fig3} the RIXS spectrum recorded at $Q=(0.35,0.35)\pi$, together with our phenomenological model. The fit includes a quadratic polynomial representing the RIXS fluorescence, and a Gaussian lineshape at zero energy transfer to approximate quasielastic scattering.

\begin{figure}
\includegraphics[width=0.9\columnwidth,trim= 0pt 8pt 0pt 0pt, clip]{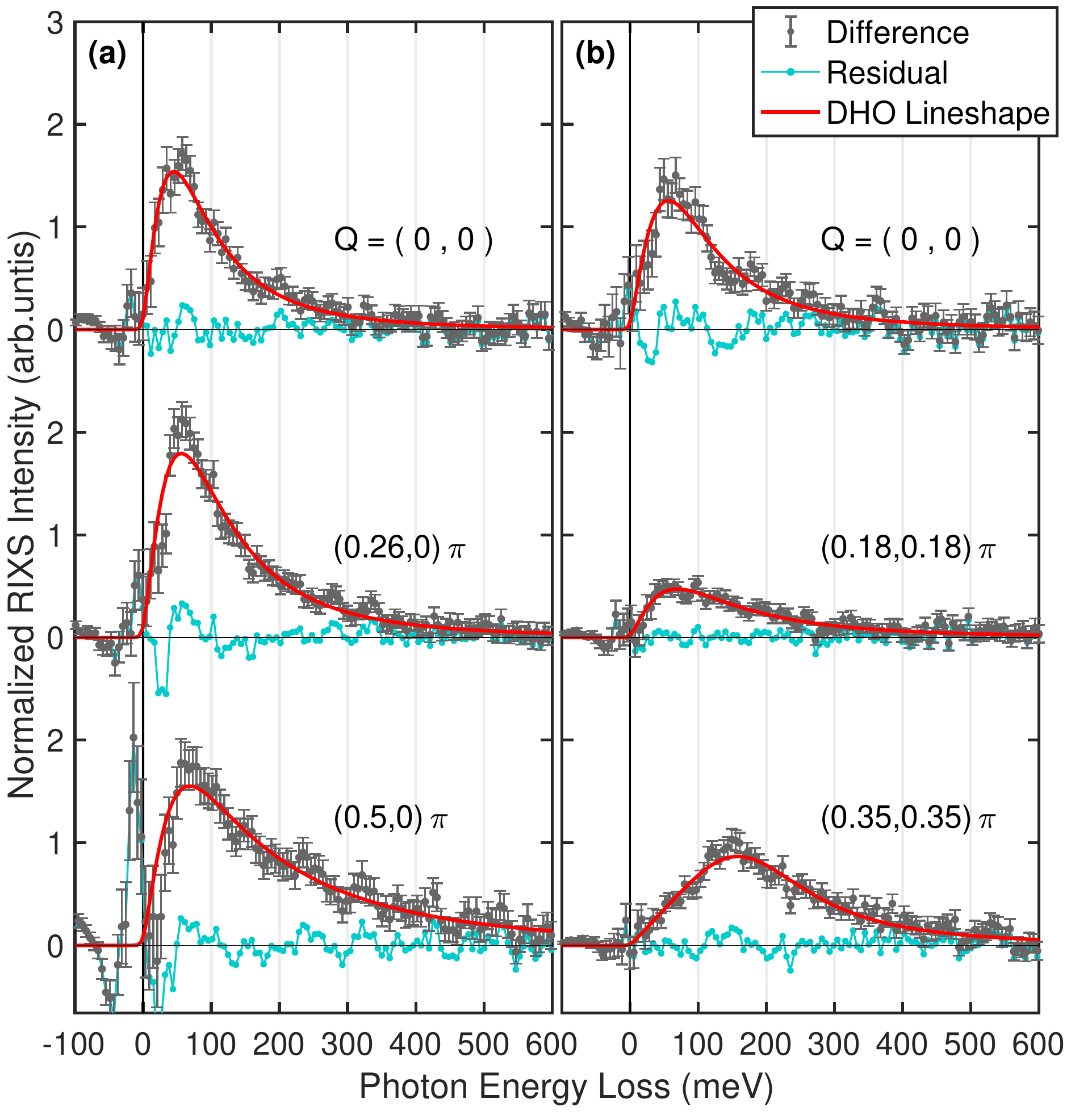}
\caption{\label{Fig4} Difference signals after subtraction of various fitted components [cf. Fig.~\ref{Fig3}(b)] from the raw data. Red lines indicate the approximation by a damped harmonic oscillator lineshape. Panels (a) and (b) illustrate representative datasets for momentum transfers along the (1,0) and (1,1) directions, respectively.}
\end{figure}

Subtraction of these features reveals the anomalous intensity shown in Fig.~\ref{Fig4}. As a simple description of these peaks, we use the dynamical structure factor $S(Q,E)$ of a damped harmonic oscillator (DHO). This model is frequently used to represent the RIXS response of paramagnon excitations~\cite{Monney2016}. We adopt the expression by Lamsal and Montfrooij, which encompasses the case of over-damping~\cite{Lamsal2016}:
\begin{equation}
S(Q,E)=A\,\frac{E}{1-{\rm e}^{-\beta E}}\frac{2\,\gamma\,E_0}{\left(E^2-E_0^2\right)^2+(E\,\gamma)^2}.\label{Eq1}
\end{equation}
Here, $Q$ and $E$ are the momentum and energy transferred to the sample, $A(Q)$ is a momentum-dependent scale factor, $\beta=1/k_\mathrm{B}T$ is the inverse Boltzmann temperature factor, and $\gamma(Q)$ and $E_0(Q)$ are the momentum-dependent damping constant and the natural (undamped) energy of the fluctuation. The excitation becomes over-damped (i.e. decays exponentially, without oscillation) if $E_0<\gamma/2$.

\begin{figure*}
\includegraphics[width=2\columnwidth,trim= 0pt 5pt 0pt 0pt, clip]{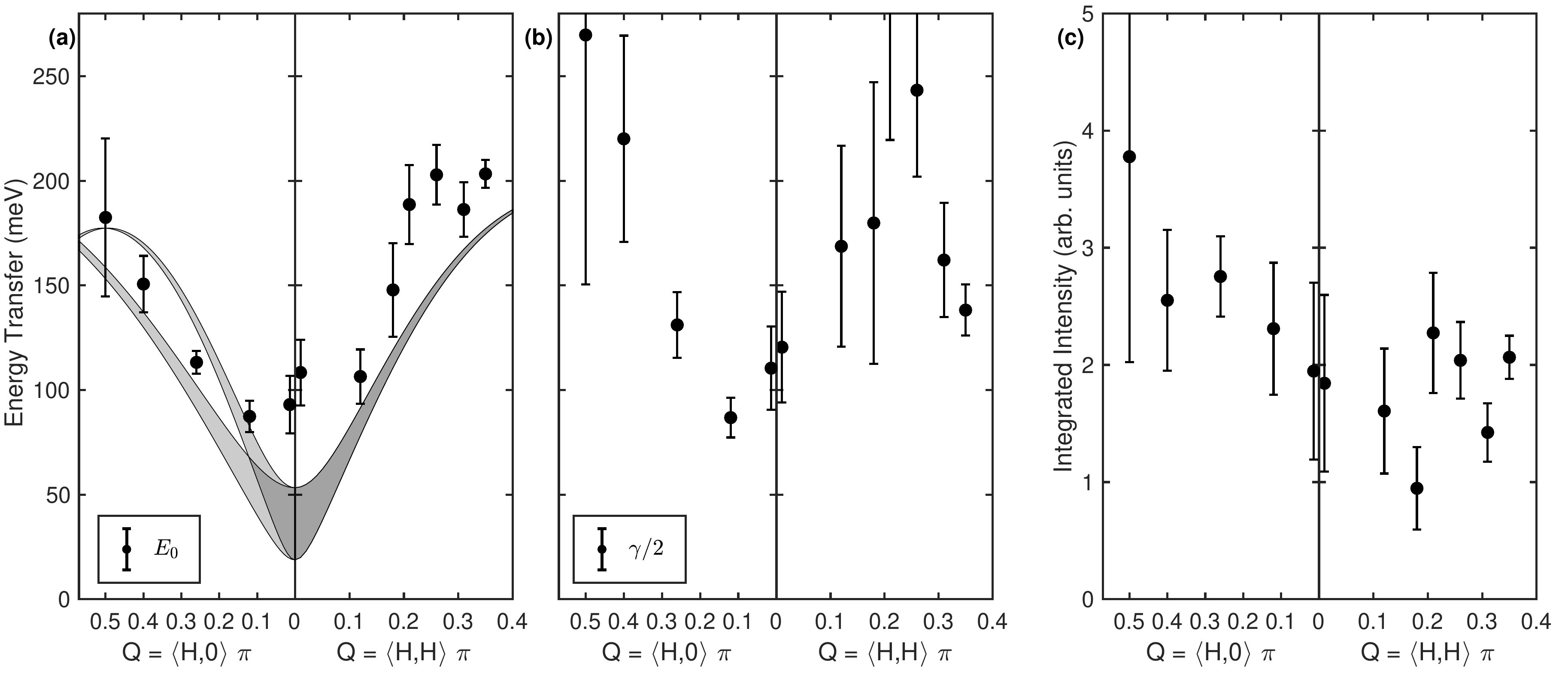}
\caption{\label{Fig5} Characteristics of low-energy RIXS excitations in FeSe, modeled by a damped harmonic oscillator lineshape, as shown in Fig.~\ref{Fig4}. Error bars indicate the standard deviation from the best fit parameters of a simultaneous least squares fit of all contributions to the spectra. They may appear uncharacteristically large in cases where features overlap and fit parameters become correlated. This becomes more evident in the overview of all fitted datasets provided in the Supplemental Material~\cite{supplemental}. The three panels show (a) natural energies $E_0$, (b) damping parameters and (c) the integrated intensity of the DHO lineshape. For comparison to a better known compound, the linear spin wave dispersion inferred from INS measurements on BaFe$_2$As$_2$~\cite{Harriger2011} is superimposed as a shaded curve in panel (a). The shading indicates the bandwidth of the dispersion in the out-of-plane momentum transfer direction, and two curves along $\langle H,0\rangle$ show the distinct dispersions along the inequivalent $a$ and $b$ directions in a detwinned sample.}
\end{figure*}

Least squares fits of datasets representative for the dispersion along the (1,1) and (1,0) directions of reciprocal space are shown in Fig.~\ref{Fig4}(a) and (b), respectively. We find that fits of all contributions discussed above generally converge without the the need to impose constraints on $A$, $E_0$ or $\gamma$. The dispersion of these fit parameters is presented in Fig.~\ref{Fig5}. Error bars, indicating the standard deviation from the best fit parameters, may appear uncharacteristically large where several scattering contributions overlap and the fit parameters are correlated. This becomes more evident in the overview of all fitted data provided in the Supplemental Material~\cite{supplemental}.

The resonance energy $E_0$ shows a V-shaped dispersion up to $\approx200\,$meV, with a $\approx100\,$meV gap at the Brillouin zone center. We detect no strong systematic variation of the integrated intensities. The fitted values of $\gamma$ are significantly larger than the experimental energy resolution (55\,meV) and appear to be significantly stronger for fluctuations along $\langle H,0\rangle$ compared to $\langle H,H\rangle$. The anisotropy  of the spin dispersion of BaFe$_2$As$_2$ in Fig.~\ref{Fig5}(a) suggests that this could be due to the orthorhombic twinning of our sample. For all spectra, the DHO lineshape appears strongly damped or overdamped, which is closely reminiscent of RIXS excitations observed in La$_{1.77}$Sr$_{0.23}$CuO$_4$~\cite{Monney2016}. This is evidence for the itinerant nature of the magnetic excitations (see e.g. the neutron spectra of SrFe$_2$As$_2$, which also show a significant broadening~\cite{EWI12}).

Our analysis succeeds in describing the data with a minimal number of fitting parameters. We emphasize, however, that the measured signal is small and the background complex, and there may well exist other methods to model the data. For example, should phonon scattering be present at a significant level, then it could decrease the natural energy and increase the damping inferred from our fit of an effective DHO lineshape.

\section{Discussion}\label{Discussion}

For FeSe, as a paramagnetic metal retaining only limited local degrees of freedom, the RIXS response may include not only lattice and magnetic excitations, but also non-spin-flip $d$--$d$ fluctuations within the five-orbital manifold, as well as Fermi liquid excitations~\cite{Schuelke2007,Ament2011}. Our Bethe-Salpeter calculations demonstrate that additional spectral weight peaked in the region of 100--200\,meV is unlikely to originate from localized inter-orbital ($d$--$d$) excitations. Similarly, for reasonable electron-phonon coupling, we are able to rule out dominant phonon scattering in this energy-range. This leaves propagating magnetic fluctuations as the most likely explanation for this signal.

If the response in the present RIXS channels ($\pi\pi'$+$\pi\sigma'$) is dominated by spin-flip processes, a correspondence with the dynamic magnetic susceptibility measured by INS would be expected. Inelastic neutron studies of FeSe have revealed a spectrum of anisotropic paramagnon excitations~\cite{Rahn2015,Wang2016,Shamoto2015} steeply dispersing out of $(\pi,0)$, with a band width of about 200\,meV, consistent with the RIXS signal.

Due to kinematic constraints, the narrow range of reciprocal space studied by soft RIXS is not directly accessible to inelastic neutron scattering. However, INS can probe equivalent regimes around the $\Gamma$ points in neighboring Brillouin zones, where the measured paramagnon response should only differ by the reduction in the Fe magnetic form factor. One single crystal INS study has been reported in which Brillouin zone centers were accessed in the energy range of 45--160\,meV~\cite{Wang2016}. In these measurements a suppression of the magnetic scattering was observed at the Brillouin zone center. 

Similarly, inelastic neutron studies of the more strongly correlated ``122'' family of IBSCs have found a particularly strong damping of spin fluctuations close to the Brillouin zone center~\cite{Harriger2011}. In this case, a comparison is possible to the high-resolution Fe $L_3$ edge RIXS study of (Ba,K)Fe$_2$As$_2$ reported by Zhou \etal~\cite{Zhou2013}. The authors demonstrated a close agreement with the spin-wave model which had been used to model INS spectra, thus providing strong evidence that the RIXS signal is largely due to magnon excitations. By contrast, here we have studied a significantly more itinerant material, with no magnetically-ordered phase available as a reference. As a comparison, in Fig.~\ref{Fig5}(a) we overlay the linear spin wave model inferred for BaFe$_2$As$_2$~\cite{Harriger2011} onto the dispersion of the DHO-like lineshape in FeSe.

Several theoretical studies of magnetic fluctuations in FeSe have been reported. Based on density functional calculations, Essenberger \etal~studied the evolution of spin fluctuations as a function of the Se-height parameter $z_\mathrm{Se}$~\cite{ESS12}. While the authors simulated paramagnon spectra only for hypothetical values of $z_\mathrm{Se}$, a common result appears to be that the paramagnon dispersion is strongly gapped at the Brillouin zone center. Starting from a phenomenological tight-binding model, Kreisel~\etal~have also found paramagon branches steeply dispersing from a minimum at $(\pi,0)$~\cite{Kreisel2015}. However, the characteristics of this dispersion at the Brillouin zone center is unclear, since the calculated dynamic magnetic susceptibility appears to be almost entirely suppressed within $\approx$ 0.25\,r.l.u. of the $\Gamma$ point.

Kaneshita \textit{et al.} have calculated Fe $L_3$-edge RIXS spectra based on a five-band Hubbard model of metallic IBSCs~\cite{Kaneshita2011}. In contrast to cuprates, the authors find that the magnon intensities measured in RIXS are not well separated from $d$--$d$ excitations. Instead, the spectra are dominated by orbital excitations that are not associated with a spin-flip. However, the electronic correlations of $U=1.2$\,eV assumed in these calculations are likely an underestimation in the case of FeSe, where values around $U\approx 4$\,eV are appropriate~\cite{Aichhorn2010}. Being a description of the itinerant limit in IBSCs, the result by Kaneshita~\etal~may therefore not be applicable to the present data.

\section{Conclusions}
In conclusion, we have reported a resonant x-ray spectroscopic study at the Fe $L_3$ edge of the unusual superconductor FeSe, with a combined energy resolution of 55\,meV FWHM. The key challenge in this context is the extreme weakness of local excitations in a material where most RIXS spectral weight is due to excitations in the Fermi liquid continuum. We observe significant structure in the low-energy regime, which can likely be associated with magnetic fluctuations.

The excitation energies obtained from fitting a DHO-lineshape to the RIXS spectra reveal a V-shaped dispersion between 100 and 200\,meV, with a large spin-gap at the Brillouin zone center. Since no neutron inelastic study of FeSe is available which has probed the relevant momentum- and energy--transfer regime, and linear spin wave models are found to be inadequate in this regime~\cite{Wang2016}, the identification of these excitations with the dynamic magnetic susceptibility of FeSe is not straightforward. Nevertheless, our calculations of the RIXS cross section, based on an LDA model of FeSe and knowledge of the principle phonon modes, rule out orbital ($d$--$d$) or lattice excitations as a possible origin. Instead, the observed dispersion is qualitatively similar to that observed in parent compounds of FeAs-based superconductors.

For correlated itinerant states as in FeSe, the energy resolution of RIXS is presently limited by the measurable intensity. The realization of this technique's full potential, e.g.~including full polarization analysis to distinguish between the orbital character of the excitations, will have to await further advances in synchrotron and beamline instrumentation. On the other hand, it would be interesting to extend the present dataset by investigating the resonant behavior (incident energy dependence) of the spectra. It would also be relevant to obtain corresponding measurements above the tetragonal-to-orthorhombic phase transition. We hope that the quantitative interpretation of such results will be aided by increasingly sophisticated \textit{ab initio} calculations of the RIXS response due to phonons, spin-flip and non-spin-flip orbital excitations.
~\\
\begin{acknowledgments}
We are grateful to the European Synchrotron Radiation Facility for provision of beamtime at the ID32-ERIXS endstation. This work was supported by the U.K. Engineering and Physical Sciences Research Council (grant no.~EP/J017124/1). MCR is grateful for support through the Oxford University Clarendon Fund, the LANL Director's Fund and the Humboldt Foundation. We thank Andrey Geondzhian for assistance with estimating the phonon contribution to the RIXS spectrum.
\end{acknowledgments}

\balance
\putbib[FeSeBib]

\end{bibunit}

\clearpage

\cleardoublepage
\onecolumngrid
\appendix

\begin{center}

\Large
Supplemental Material:

\vspace{0.5cm}

\large
{\bf Paramagnon dispersion in $\beta$-FeSe observed by Fe $L$-edge resonant inelastic x-ray scattering}

\vspace{0.5cm}

\normalsize

M. C. Rahn,$^1$ K. Kummer,$^2$ N. B. Brookes,$^2$ A. A. Haghighirad,$^1$ K. Gilmore,$^2$ and A. T. Boothroyd$^1$

\vspace{0.2cm}
\small

$^1${\it Department of Physics, University of Oxford, Clarendon Laboratory, Oxford, OX1 3PU, United Kingdom}\\[1pt]
$^2${\it European Synchrotron Radiation Facility, Bo\^{i}te Postale 220, F-38043 Grenoble Cedex, France}\\[1pt]

\end{center}
\vspace{1cm}

\begin{bibunit}[apsrev4-1]

\begin{center}
{\bf 1. Laboratory single crystal x-ray diffraction}
\end{center}
 In order to select samples of best crystalline quality, a number of $\beta$-FeSe single crystals were characterized by room temperature four-circle Mo $K_\alpha$ x-ray diffraction (Agilent Supernova). The crystals generally grow as rectangular platelets and x-ray diffraction revealed that their edges are parallel to the tetragonal unit cell (space group $P4/nmm$, with lattice parameters $a=3.77$\,\AA, $b=5.52$\,\AA). Figure S1~(b) shows an example of such laboratory x-ray data, where scattered intensity is integrated over a margin perpendicular to the $(H,K,0)$ plane of reciprocal space. The in-plane mosaicity of the observed Bragg reflections is smaller than the instrumental resolution ($\approx0.6^\circ$). Since the material is relatively soft, platelets easily bend, which causes imperfections in the vertical stacking of the atomic planes. Accordingly, crystals with the smallest mosaic spread along the $(0,0,L)$ direction were selected for RIXS measurements.

\begin{figure}[H]
\centering
\includegraphics[width=0.6\columnwidth,trim= 0pt 0pt 0pt 0pt, clip]{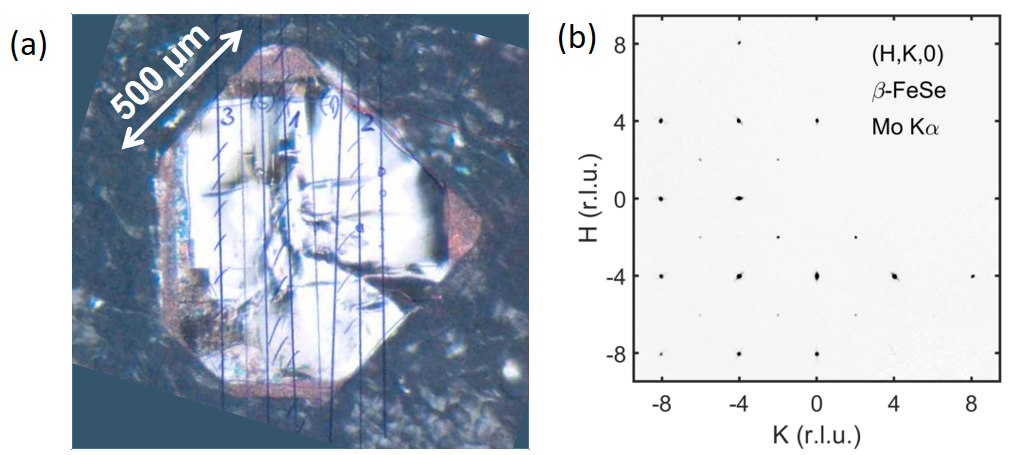}%
\\[2pt]
\justify{\label{FigS1} FIG.~S1 (a) Micrograph of a $\beta$-FeSe single crystal. The oxidized surface has been stripped from the platelet using adhesive tape.  (b) (H,K,0) intensity map of reciprocal space, created from Mo $K_\alpha$ four-cricle x-ray diffraction data (indexed in space group $P4/nmm$).}
\end{figure}

\newpage

\begin{center}
{\bf 2. Magnetization}
\end{center}
For selected crystallites, we performed magnetization measurements using a superconducting quantum interference device (SQUID, Quantum Design). The zero-field-cooled temperature sweep shown in Fig.\,S2 illustrates the clean onset of ideal diamagnetism at the superconducting critical temperature $T_c=8.9$\,K.

\begin{figure}[H]
\centering
\includegraphics[width=0.5\columnwidth,trim= 0pt 0pt 0pt 0pt, clip]{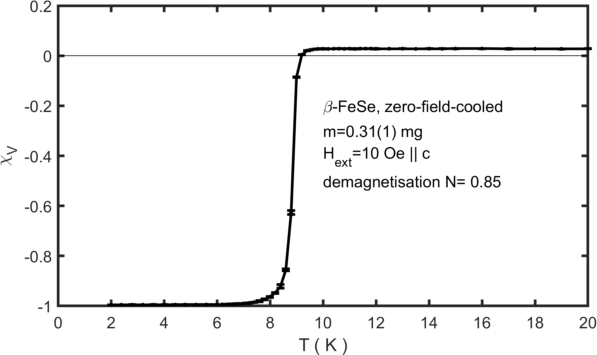}%
\\[2pt]
\justify{\label{FigS2} FIG.~S2. SQUID magnetometry of a 0.31(1)\,mg $\beta$-FeSe platelet that was used in the RIXS study. The crystal was cooled in zero-field before applying an excitation field of 10\,Oe along the crystallographic $c$ axis. The magnetic susceptibility was then measured upon warming, which reveals the superconducting transition at $T_c=8.9$\,K. The data shown here are normalized under the assumption of a 100\% superconducting volume fraction (demagnetization factor $N=0.85$).}
\end{figure}

\newpage
\vspace{2cm}
\begin{center}
{\bf 3. Data collection and processing}
\end{center}

Due to the thermal drifts of the monochromator, the position of the image on the CCD chip shifts over time. Data were therefore collected in short cycles of around five minutes. To obtain statistics as shown in the main text, spectra corresponding to a total measurement time of ca. 7--8 hours were summed for each spectrum.

Using the RixsToolBox application, see Ref.~[\onlinecite{Kummer2017}], all scans comprising one measurement were calibrated and centered with respect to each other by fast Fourier transform and convolution of the spectra. This process did not significally worsen the effective energy resolution compared to individual datasets, which confirms the reliability of the algorithm. All spectra were obtained in a scattering geometry in which the quasielastic line is observed and a Gaussian peak can be fitted at this position. This serves as a reference of zero energy transfer, $E=0$. The amount of thermal drift over the minimal exposure time required to obtain adequate counting statistics on the quasielastic line is one limiting factor to the energy resolution of the experiment. 

Several of the spectra were obtained whilst avoiding to irradiate any position on the sample surface for more than five minutes. After each irradiation, the TEY spectra were confirmed to be fully of Fe$^{2+}$ character, before slightly translating the sample. Since the vertical size of the beam profile at the sample is very small ($\approx4\,\mu$m), this procedure is feasible even for small sample dimensions.

\begin{figure}[H]
\centering
\includegraphics[width=0.5\columnwidth,trim= 0pt 0pt 0pt 0pt, clip]{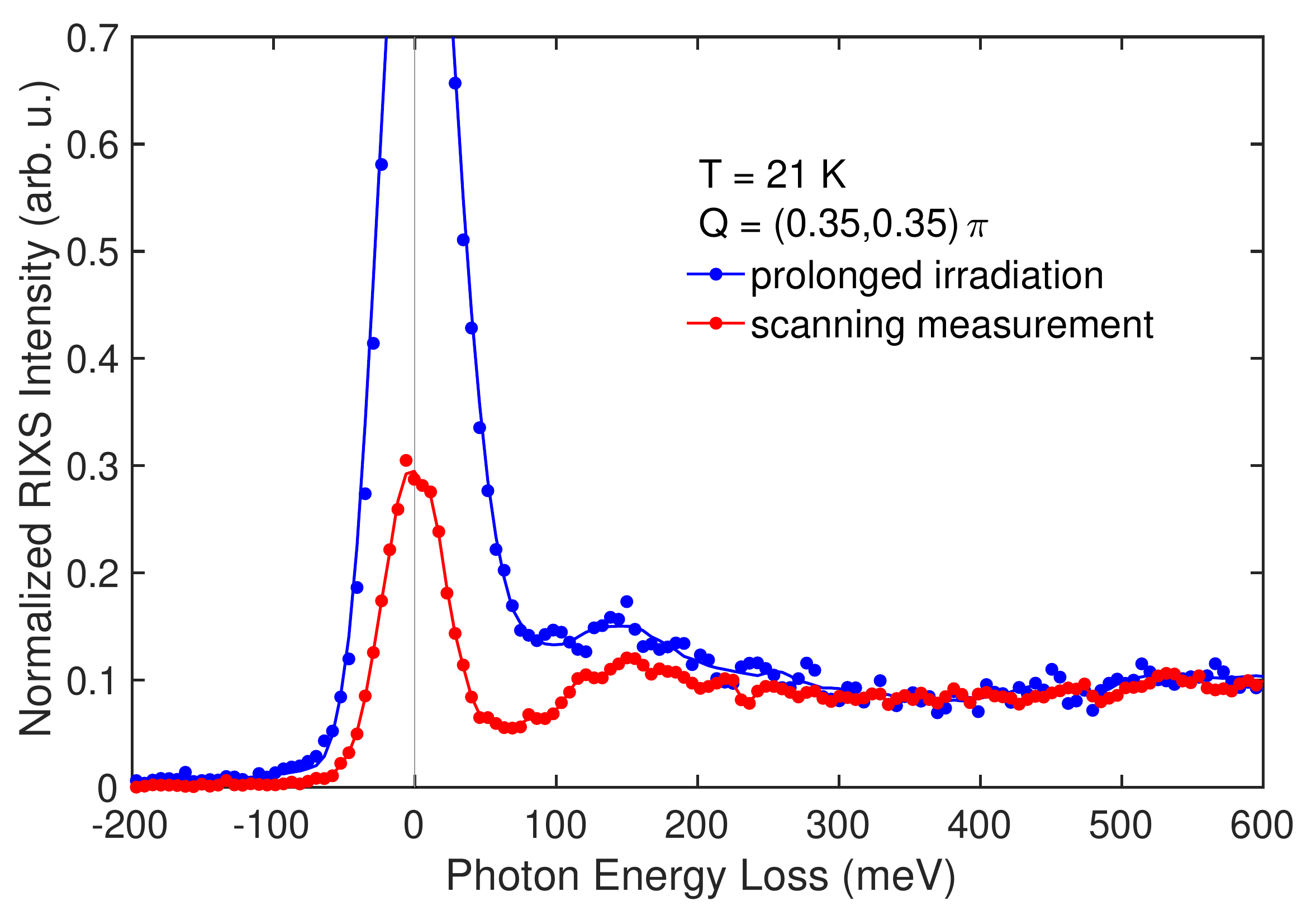}%
\justify{\label{FigS3} FIG.~S3. Comparison of RIXS spectra obtained at the same momentum transfer, incident angle and incident energy, after sustained irradiation of the same spot on the sample surface (blue) and when translating the sample position every five minutes (red). A strong enhancement of the elastic peak appears to be the primary effect of the surface damage.}
\end{figure}

In Fig.~S3 we compare spectra obtained in this \textit{scanning mode} and under continuous irradiaton, with otherwise identical conditions. It appears that the main effect of irradiation is a growth of the elastic line in the RIXS spectra. This could be qualitatively explained by the fact that the incident energy (707\,eV) lies slightly below the $L_3$ absorption edge for trivalent iron. Since the pre-edge regime is dominated by strong elastic scattering, this would explain the enhancement of the $E=0$ peak as iron atoms in the surface layers are being oxidized. The additional RIXS spectral weight centered around 150--300\,meV does not appear to be altered. We therefore assume that this is a bulk response of FeSe which is not strongly affected by irradiation.

\newpage
\begin{center}
{\bf 4. RIXS fluorescence and phonon calculations}
\end{center}
The results of density functional theory (DFT) based Bethe Salpeter equation (BSE) calculations of the fluorescence contribution to the RIXS spectrum are shown in Fig.~S4(a)--(b).  Calculations assumed the experimental crystal structure.  The DFT portion of the spectral calculations were performed with the Quantum-ESPRESSO code~\cite{Gianozzi2009}, which employs pseudopotentials and periodic boundary conditions.  We used LDA norm-conserving pseudopotentials with 6 electrons in valence for Se and 16 electrons in valence for Fe.  Convergence required a planewave basis energy cutoff of 120\,Ry for the wavefunctions and 480\,Ry for the charge density.  The ground-state charge density was converged with a $7\times7\times5$ $k$-point sampling while $9\times9\times7$ $k$-points were used for the BSE final states.  Numerical, full frequency calculation of the $G_0W_0$ self energy was performed with the Abinit code~\cite{Gonze2002,Gonze2016,Bruneval2006}.

To estimate the phonon contribution to the RIXS spectra we evaluated a simple model for the coupling of local electronic levels to local vibrational modes~\cite{Ament2011a}.  We considered two electronic levels coupled to two vibrational modes of energies 25\,meV and 40\,meV.  The Fe 2$p_{3/2}$ core-hole lifetime was set to the known value of 180\,meV HWHM linewidth~\cite{Krause1979}.  For fairly strong values of the coupling constants $g=(M/\omega_{ph})^2=5$ ($\omega_{ph}$ is the phonon frequency and $M$ the electron-phonon coupling energy), we find that the phonon RIXS contribution peaks around 40\,meV with a rapidly decaying tail to higher energy as shown in Fig.~S4(c).  Significantly increasing the coupling constants does not shift the peak position, but only causes the high energy tail to decay somewhat more gradually.  That the phonon contribution peaks at the first harmonic is a largely inescapable consequence of the relatively short core-hole lifetime of the Fe 2$p_{3/2}$ level and it appears highly unlikely the experimental intensity around 150\,meV (equivalent to a third or fourth phonon harmonic) would be vibrational in origin.

\begin{figure}[H]
\centering
\includegraphics[width=0.96\columnwidth,trim= 0pt 0pt 0pt 0pt, clip]{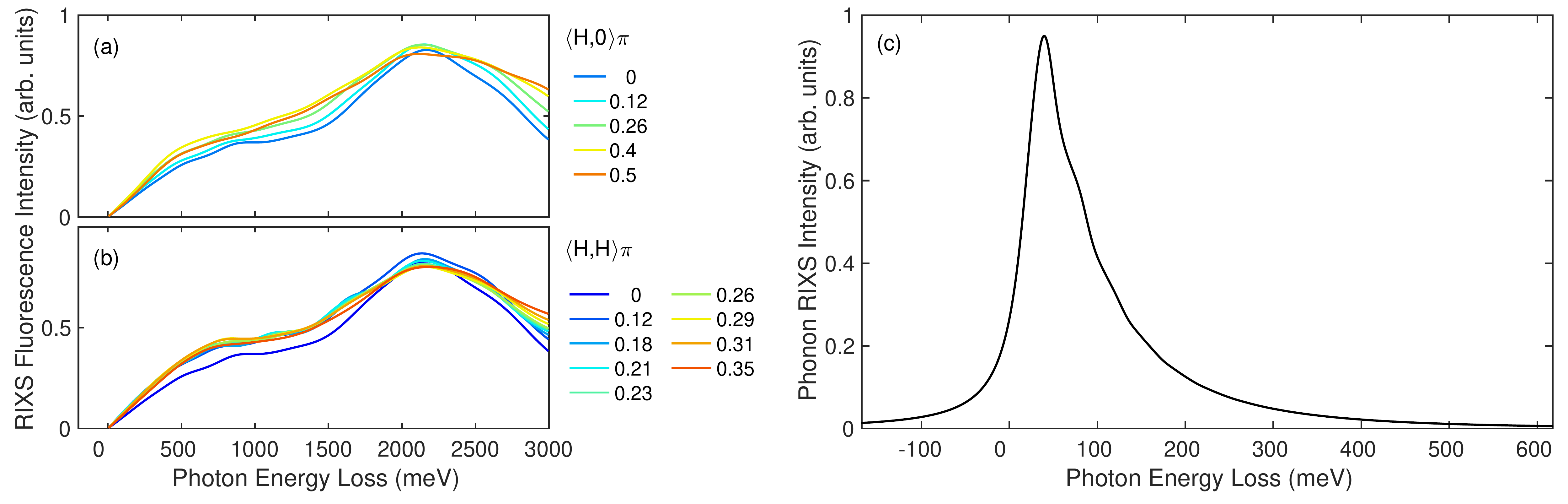}%
\justify{\label{FigS4} FIG.~S4. (a) Electronic quasiparticle contribution to the RIXS spectrum of FeSe, obtained through evaluation of the Kramers-Heisenberg equation within the framework of the many-body Bethe-Salpeter equation. (b) Calculated RIXS phonon spectrum of FeSe, assuming a core-hole lifetime broadening of 180\,meV and principle phonon modes at 20.5, 25.5 and 40\,meV, as observed in the Raman study by Zakeri\,\textit{et al.}~\cite{Zakeri2017}}.
\end{figure}

\newpage
\begin{center}
{\bf 5. Summary of RIXS spectra}
\end{center}

All RIXS spectra used to construct the dispersion shown in Fig.\,5 of the manuscript are summarized in Figs.~S5 and S6 for the (1,0) and (1,1) directions of reciprocal space, respectively. Subpanels labeled (b) and (c) indicate phenomenological fits of the contributions discussed in the main text, in analogy to Fig.~3 therein.

\begin{figure}[H]
\includegraphics[width=1\columnwidth,trim= 0pt 0pt 0pt 0pt, clip]{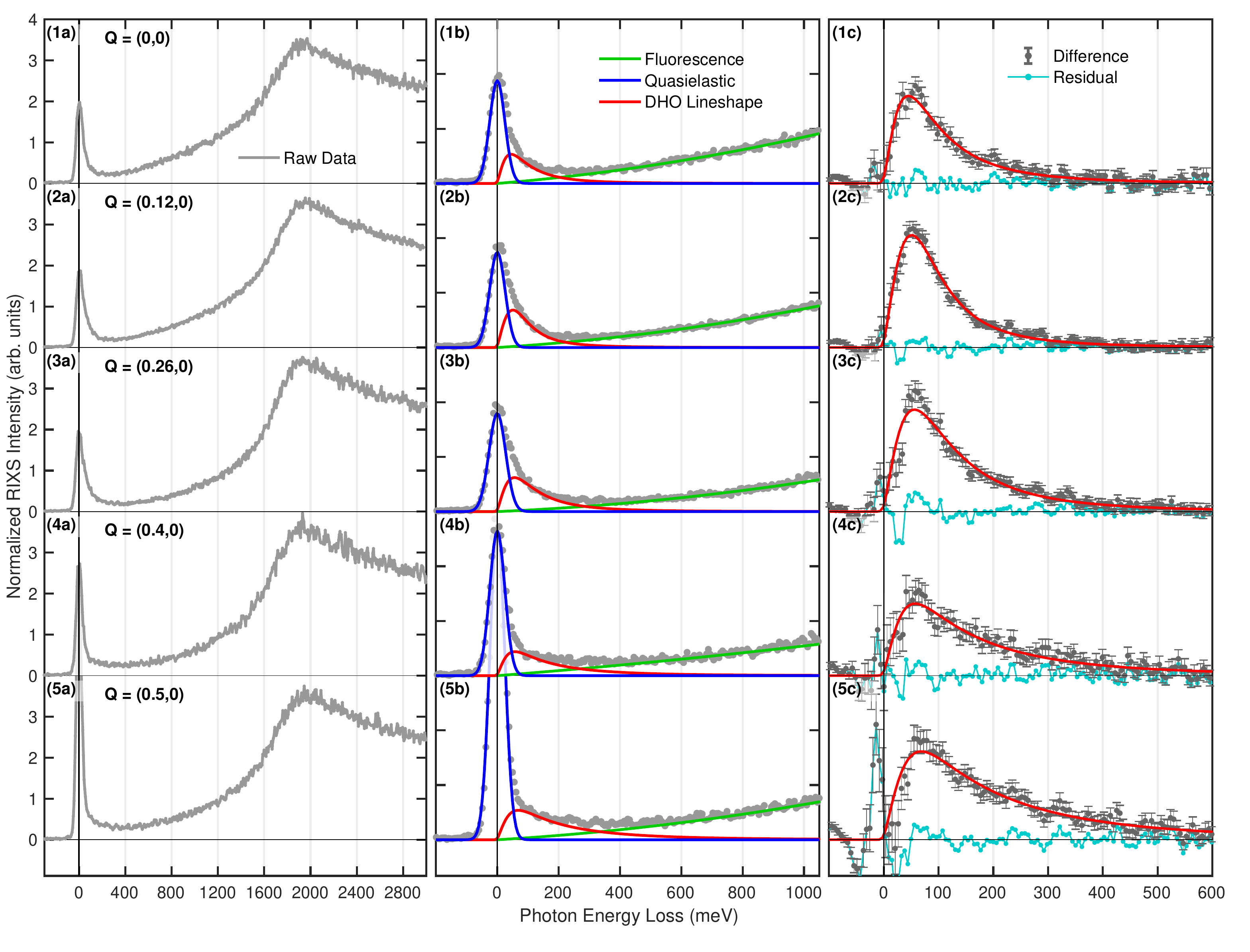}
\justify{\label{FigS5} FIG.~S5. Fe $L_3$ RIXS spectra of $\beta$-FeSe recorded at 21\,K, with an in-plane momentum transfer along the $(1,0)$ direction of reciprocal space. (1a)--(5a) Raw data, dominated by quasielastic scattering and a strong fluorescence contribution. (1b)--(5b) Detailed views of the low-energy region, with curves indicating a fit of three contributions to the signal. (1c)--(5c) Difference signals after subtraction of the quasielastic line and fluorescence contribution (grey errorbars). Residual after additional subtraction of the DHO lineshape (cyan markers).}
\end{figure}
\newpage

\begin{figure}[H]
\centering
\includegraphics[width=0.96\columnwidth,trim= 0pt 0pt 0pt 0pt, clip]{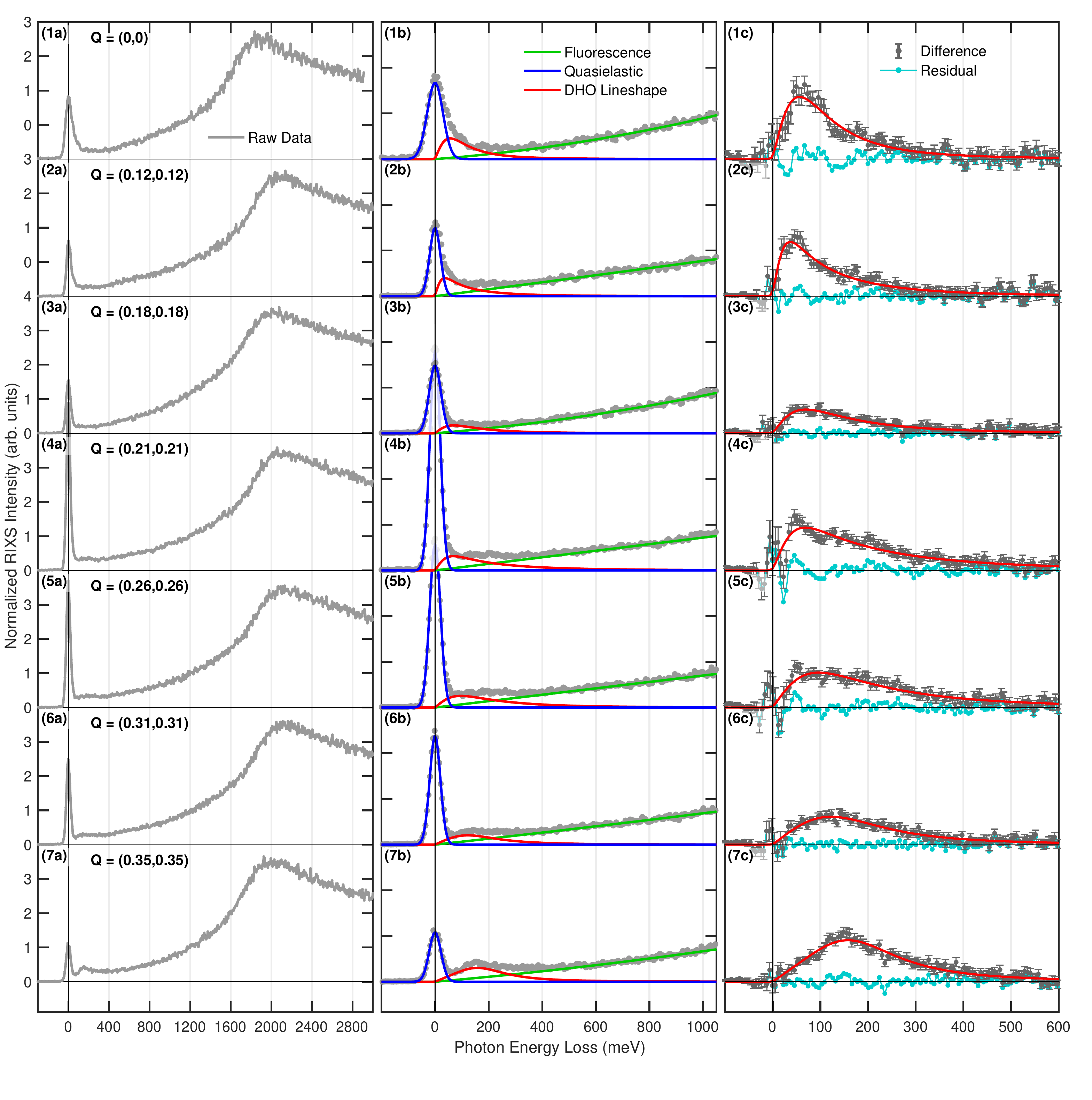}
\justify{\label{FigS6} FIG.~S6. Fe $L_3$ RIXS spectra of FeSe recorded at 21\,K, with an in-plane momentum transfer along the $(1,1)$ direction of reciprocal space. The data are presented in analogy to Fig.\,S4.}
\end{figure}

{\LARGE
\putbib[FeSeBib]}
\end{bibunit}

\end{document}